\documentclass[12pt,a4paper]{article}
\usepackage{amsmath,amsfonts,epsfig}
\usepackage{amssymb,bbold}
\usepackage{mathrsfs}
\usepackage{hyperref}
\newcommand{\tom}{\tilde{\omega}}
\newcommand{\om}{{\omega}}

\begin{document}
\numberwithin{equation}{section}
\setlength{\unitlength}{.8mm}

\begin{titlepage} 
\vspace*{0.5cm}
\begin{center}
{\Large\bf Thermodynamics in the Sine-Gordon model: the NLIE approach}
\end{center}
\vspace{1.5cm}
\begin{center}
{\large \'Arp\'ad Heged\H us}
\end{center}
\bigskip

\vspace{0.1cm}

\begin{center}
HUN-REN Wigner Research Centre for Physics,\\
H-1525 Budapest 114, P.O.B. 49, Hungary\\ 
\end{center}
\vspace{1.5cm}
\begin{abstract}

\end{abstract}
In this paper we derive Kl\"umper-Batchelor-Pearce-Destri-de Vega type nonlinear integral equations for describing 
the thermodynamics in the sine-Gordon model, when a chemical potential coupled to the topological charge 
 is also present in the theory. The equations are valid at any value of the coupling constant 
and are particularly useful for computing the expectation values of local operators and some currents 
as functions of the temperature and the chemical potential. The equations are also extended to the case, 
when an extra term corresponding to the momentum conservation, is added to the thermodynamic potential. 

The benefits of this description are twofold. On the one hand, it  can serve as an appropriate testing ground
for other promising theoretical methods, like the method of random surfaces. 
On the other hand, the efficient computation of  vertex operator expectation values, allows one to provide useful theoretical data, 
for the experimentally measurable coherence factors of a tunnel-coupled cold atomic condensate system.
\end{titlepage}

\section{Introduction}

In this paper, we would like to fill a gap in the literature of the integrable description of the 
sine-Gordon model. Namely, we derive the Kl\"umper-Pearce-Destri-de Vega \cite{KlumperPearce,Destri:1992qk} type nonlinear integral equations 
(NLIE), for the case when a chemical potential coupled to the topological charge 
 is also present in the theory. So far, only the case without chemical potential 
\cite{Destri:1992qk,Destri:1994bv}, was available in the 
literature. The thermodynamical description with chemical potential was available only in the Thermodynamic 
Bethe Ansatz (TBA) approach \cite{FW2}, which was recently rederived for the whole range of the coupling constant, and used for an extensive study of the thermodynamic properties of the theory \cite{Nagy:2023phz}.


Despite the availability of the TBA equations, their main drawback is that the number of components and the actual 
form of the equations depend very strongly and in a highly intricate way on the value of the coupling constant, 
whereas the key advantage of the NLIE approach is that the coupling dependence is fully encoded in the kernel, 
reducing the problem to solving an integral equation involving a single unknown function to obtain the thermodynamic 
potentials at arbitrary coupling.

The assumption, that there should exist a NLIE description for the sine-Gordon thermodynamics containing 
chemical potential comes from the analogy with the XXZ-model. There, from the quantum transfer matrix approach, 
 it is well, known, that an NLIE description of the thermodynamics with chemical potential can be given 
\cite{Klumper:1993vq,Destri:1994bv,Essler2005}. 
The derivation of the TBA for the sine-Gordon \cite{FW2,Nagy:2023phz} and the XXZ models  
\cite{Takahashi:1972zza}, are almost the same, only the 
source terms become different due to the different dispersion relations of the elementary excitations of the 
two models. Thus, one can come to the conclusion, that an NLIE being analogous to that of the XXZ model,
 can be derived for the sine-Gordon model, too. 

Actually, we will do this in this paper. We show, that the analogous NLIE for the sine-Gordon model can be 
 derived easily, and the final result is also analogous to the XXZ case. Namely, only an imaginary 
valued twist parameter should be inserted into the equations. 


Having derived the equations, we test them against some simple TBA cases, and extend them, to the case, when 
the conserved momentum is also included into the thermodynamic potential. We also discuss the computation 
of the expectation values of  conserved currents and local operators of the theory. One can state with 
confidence,  
that the main strength of the NLIE formalism is that, it offers an efficient framework for 
 computing the thermodynamic potentials and expectation values of currents and local operators at any 
values of the coupling constant of the theory. 


In the rest of the paper we use the equivalence between the sine-Gordon and massive Thirring models \cite{s-coleman,Klassen:1992eq}, 
and we will mostly work in the fermionic language, which seems to be more natural for our argumentation and computations. 

The plan of the paper is as follows. 
In section \ref{sect2}. we argue that in the language of the massive Thirring model, the introduction of a chemical potential 
is equivalent to imposing certain chemical potential dependent twisted boundary conditions on the fermion-fields 
in the direction of Euclidean time.
In section \ref{sect3}, we show how to derive the nonlinear-integral-equations for describing the free-energy density 
at any temperature and chemical potential. 
In section \ref{sect4}, we perform a few very simple tests and checks on our equations in the UV/IR regimes. 
 In section \ref{sect5}, we generalise the equations by adding the momentum as a conserved quantity to the thermodynamic 
potential, and show  how to compute expectation values of the currents and local operators in the NLIE formalism. 
We close the paper, with a short summary in section \ref{sect6}.

The single appendix \ref{appA}, contains some numerical comparison to results coming from the Thermodynamic Bethe Ansatz (TBA) equations at very specific values of the coupling constant.

\section{sine-Gordon/massive Thirring model with chemical potential} \label{sect2}

The theories, the thermodynamics of which we consider in this paper are the sine-Gordon and massive Thirring models.

The sine-Gordon model is defined by the Euclidean action as follows: 
\begin{equation}
\label{sG_Lagrangian}
{S}_{SG}[\Phi]\!=\!\!\int \! dx\, d\tau  \left\{ \displaystyle\frac{1}{2} (\partial _{\mu }\Phi(x,\tau))^2 -\displaystyle\frac{2\, \kappa^{2}}{\sin(\beta_{SG} ^{2}/8)}\cos \left( \beta_{SG} \Phi(x,\tau) \right) \right\}\,  \qquad 0<\beta_{SG}^2<8 \pi.
\end{equation}

The Euclidean action of the massive Thirring model takes the form as follows: 
\begin{equation} \label{Seucl}
\begin{split}
S_{MT}[\bar{\psi},\psi]=\int\limits \! d\tau dx \, {\big\{} 
-\bar{\psi}\big( \gamma_0 \partial_\tau+ \gamma_1 \partial_x+m \big)\psi
+\tfrac{g}{2} \, \bar{\psi}\gamma_\mu \psi \, \bar{\psi}\gamma_\mu \psi
{\big\}},
\end{split}
\end{equation}
where now $\gamma_\mu$ denotes the Euclidean $\gamma$-matrices satisfying the algebra
\begin{equation} \label{galg}
\begin{split}
{\big{\{}} \gamma_\mu,\gamma_\nu {\big{\}}}=2 \, \delta_{\mu \nu}, \qquad \mu,\nu=0,1.
\end{split}
\end{equation}
In the chiral representation, their matrix representation takes the form as follows: 
\begin{equation} \label{galak}
\begin{split}
\gamma_0=\begin{pmatrix} 0 & 1 \\ 1 & 0
\end{pmatrix}, \qquad 
\gamma_1=\begin{pmatrix} 0 & -i \\ i & 0
\end{pmatrix}, \qquad
\gamma_5=-i \, \gamma_0 \, \gamma_1=\begin{pmatrix} -1 & 0 \\ 0 & 1
\end{pmatrix}. 
\end{split}
\end{equation}.

It is known \cite{s-coleman} that the two models are equivalent if the coupling constants are related to each other, as follows:
\begin{equation} \label{gbeta}
1+\frac{g}{\pi}=\frac{4 \pi}{\beta_{SG}^2}.
\end{equation}
In the sequel, we will use the following, common in the literature \cite{Feverati:2000xa}, parameterization of the coupling constants:
\begin{equation}\label{pdef}
\begin{split}
p=\frac{\beta_{SG}^2}{8 \, \pi -\beta_{SG}^2}, \qquad \frac{\beta_{SG}^2}{4 \, \pi}=\frac{1}{1+g/\pi}=\frac{2 \,p}{p+1}, \qquad 0<p<\infty.
\end{split}
\end{equation}
In the language of this coupling constant the $p=1$ value correspond to the free fermion point of the theory, and the 
$0<p<1$ and $1<p$ regimes correspond to the attractive and repulsive regimes of the theory, respectively.

In both theories there is a conserved $U(1)$ current, which guarantees the conservation of the corresponding charge. 
The conserved current in the sine-Gordon model is a topological one:
\begin{equation}\label{jSG}
\begin{split}
j^{(SG)}_\mu(x)=\frac{\beta_{SG}}{2 \pi} \,  \epsilon_{\mu \nu} \partial_\nu \Phi(x), \qquad \epsilon_{01}=-\epsilon_{10}=1, 
\qquad \epsilon_{00}=\epsilon_{11}=0.
\end{split}
\end{equation}
which ensures the conservation of the topological charge:
\begin{equation}\label{Qtop}
\begin{split}
Q_{top}=\frac{\beta_{SG}}{2 \, \pi} \, \int\limits_{-\infty}^\infty dx \, \partial_x \, \Phi(x,\tau).
\end{split}
\end{equation}
The analogous quantities in the massive Thirring model are the fermion current, 
\begin{equation}\label{jMT}
\begin{split}
j^{(MT)}_\mu(x,\tau)=\bar{\psi }(x,\tau)\gamma_{\mu }\psi(x,\tau)
\end{split}
\end{equation}
and the corresponding conserved fermion charge:
\begin{equation}\label{Qmt}
\begin{split}
Q= \int\limits_{-\infty}^\infty dx \, \bar{\psi }(x,\tau)\gamma_{0 }\psi(x,\tau).
\end{split}
\end{equation}
These currents and charges are mapped onto each other under the equivalence of the two models.


In this paper we are interested in the thermodynamics of these models in the presence of nonzero chemical potential coupled to the 
$U(1)$ charge of the theory. We compute the free-energy density of these models in this scenario. 
\begin{equation}\label{f0}
\begin{split}
f(\beta,\mu)=\lim_{L \to \infty} -\frac{1}{L \, \beta} \, \log \mbox{Tr}(e^{-\beta\, (H_L-\mu \, Q)}),
\end{split}
\end{equation}
 where $\beta=1/T$ is the inverse temperature, $H_L$ is the Hamiltonian defined in volume $L, 
 $ $Q$ is the topological/fermion charge and $\mu$ is the chemical 
potential associated to it. 

The introduction of a chemical potential coupled to a conserved charge can be formulated in two ways in a 
quantum field theory. Either, by imposing twisted boundary conditions \cite{Fendley:1993jh} 
or by inserting a defect line on which the excitations of the theory can scatter \cite{Bajnok:2007jg,Ahn:2011xq}. 

In this paper, we find more appropriate to formulate our problem in terms of twisted boundary 
conditions. In the sequel, we will work in the language of the massive Thirring model, which proves 
to be a natural framework for our computations. 


Adding a chemical potential $\mu$ to the theory is equivalent to adding the following extra term to the action:
\begin{equation}\label{Smu}
\begin{split}
\delta S_\mu[\bar{\psi},\psi]= -\mu \, \int\limits \! d\tau dx \,  \bar{\psi}(x,\tau) \gamma_0 \psi(x,\tau).
\end{split}
\end{equation}
Adding this extra term to the action (\ref{Seucl}), the modified action can be recasted as the original action 
with a $\partial_\tau \to \partial_\tau+ \mu$ replacement, which has important consequences.
\begin{equation} \label{Seuclmu}
\begin{split}
S_\mu[\bar{\psi},\psi]=\int\limits \! d\tau dx \, {\big\{} 
-\bar{\psi}\big( \gamma_0 (\partial_\tau+ \, \mu) + \gamma_1 \partial_x+m \big)\psi
+\tfrac{g}{2} \, \bar{\psi}\gamma_\mu \psi \, \bar{\psi}\gamma_\mu \psi
{\big\}}.
\end{split}
\end{equation}
The free-energy of the model can be obtained from the partition function given by the Euclidean path-integral as follows:
\begin{equation}\label{partpath}
\begin{split}
Z(\beta,\mu)=\int\limits_{{}_{"\tau+\beta=\tau "}}\!\! 
{\cal D}\bar{\psi} {\cal D}{\psi} \, \, e^{-S_\mu[\bar{\psi},\psi]},
\end{split}
\end{equation}
where $\beta=1/T$ is the inverse temperature, and in the direction of the Euclidean time $\tau,$ 
anti-periodic boundary conditions should be imposed on the fermion fields:
\begin{equation}\label{antiper}
\begin{split}
\psi(x,\tau+\beta)=-\psi(x,\tau), \qquad \bar{\psi}(x,\tau+\beta)=-\bar{\psi}(x,\tau).
\end{split}
\end{equation}
Expanding the fermion fields in Fourier modes in the time direction, the Matsubara frequencies will 
enter the expansion:
\begin{equation}\label{tFFT}
\begin{split}
\psi(x,\tau)=\sum\limits_{n \in \mathbb{Z}}\, e^{i \, \omega_n \, \tau} \, \psi_n(x), \qquad \bar{\psi}(x,\tau)=\sum\limits_{n \in \mathbb{Z}} \, e^{-i \, \omega_n \, \tau} \, \bar{\psi}_n(x),
\end{split}
\end{equation}
where due to the anti-periodic boundary conditions the frequencies take the form:
 \begin{equation}\label{Matsub}
\begin{split}
\omega_n=\frac{2 \pi}{\beta}(n+1/2),  \qquad n \in \mathbb{Z}.
\end{split}
\end{equation}
Rephrasing the action (\ref{Seuclmu}) in terms of Fourier components of the fields, one can reinterpret 
the $\partial_\tau+ \mu$ term as if the Matsubara frequencies were shifted by $-i \, \mu.$ This shift in the 
frequencies can be interpreted as if twisted boundary conditions were imposed on fermion fields in the time direction:
\begin{equation}\label{twistedBC}
\begin{split}
\psi(x,\tau+\beta)=-e^{\beta \, \mu} \, \psi(x,\tau), \qquad 
\bar{\psi}(x,\tau+\beta)=-e^{-\beta \, \mu} \, \bar{\psi}(x,\tau),
\end{split}
\end{equation}
but the form of the usual action of the massive Thirring model (\ref{Seucl}) remained intact. 
Thus, we have shown, that the presence of the chemical potential in the theory, is equivalent to imposing 
twisted boundary conditions (\ref{twistedBC}) on the fermion fields and leave the form of the action 
unchanged.

The massive Thirring model is invariant under the exchange of the coordinates: $x \leftrightarrow \tau.$
 This allows one to apply the so-called Euclidean trick \cite{Zamolodchikov:1989cf}, 
to relate the ground state energy of the twisted model in finite 
volume $\beta,$ to the free-energy density of the original theory. 

The free energy density:
\begin{equation}\label{f}
\begin{split}
f(\beta,\mu)=\lim_{L \to \infty} -\frac{1}{L \, \beta} \, \log Z(L, \beta|\mu),
\end{split}
\end{equation}
 can be computed from two equivalent formulations of the partition function $Z(L, \beta|\mu).$ 
The first is the usual one, which forms the base of the Thermodynamic Bethe Ansatz approach:
\begin{equation}\label{Z1}
\begin{split}
Z(L, \beta|\mu)=\mbox{Tr}(e^{-\beta\, (H_L-\mu \, Q)}),
\end{split}
\end{equation}
 where $H_L$ is the Hamiltonian defined in volume $L,$ $Q$ is the fermion charge and $\mu$ is the chemical 
potential associated to it. 

Another representation comes from the coordinate exchange symmetry of the model: 
\begin{equation}\label{Z2}
\begin{split}
Z(L, \beta|\mu)=\mbox{Tr}(e^{-L \, H_{\beta,\mu}}),
\end{split}
\end{equation}
where $H_{\beta,\mu}$ is the Hamiltonian of the model in finite volume $\beta$ and with the twisted boundary 
conditions given in (\ref{twistedBC}). 

Inserting (\ref{Z2}) into (\ref{f}), and exploiting that the model is gapped, one ends up with the formula for the 
free-energy density:
\begin{equation}\label{f1}
\begin{split}
f(\beta,\mu)=\frac{E_0^{(\beta,\mu)}}{\beta},
\end{split}
\end{equation}
where $E_0^{(\beta,\mu)}$ is the ground state energy of the  massive Thirring model in finite volume $\beta$ 
with twisted periodic boundary condition (\ref{twistedBC}). 

This formula is the starting point of the derivation of the NLIE description of the Thermodynamics in the 
massive Thirring/sine-Gordon-model.


\section{The nonlinear-integral equations} \label{sect3}

The light-cone lattice approach of \cite{ddvlc}, allows one to derive nonlinear integral equations for the ground state energy of the massive Thirring model with twisted periodic boundary conditions.
The introduction of real twist parameters allowed to describe a series of interesting models, like 
the  k-folded sine-Gordon model \cite{Bajnok:2000wm}, or minimal conformal field theories 
perturbed by their relevant operator $\Phi_{1,3}$ \cite{Fioravanti:1996rz,Feverati:1999sr}. 

In our case, we will need to introduce imaginary twist parameters. However, this subtlety does not affect the 
derivation of the equations, existing results from the literature can be applied to our setup without 
modification. Following a straightforward calculation, as detailed in \cite{Destri:1994bv,Feverati:2000xa}, 
we arrive at the following equations for the continuum theory: 
\begin{equation} \label{DDVtwist}
\begin{split}
Z(\theta)&={\cal M} \beta \, \sinh \theta +\,\omega \, \frac{p+1}{p} +
\, \int\limits_{-\infty}^{\infty} \frac{d\theta'}{2 \pi i} \, G(\theta-\theta'-i\eta_+) \, L_+(\theta'+i \eta_+)- \\
&-\, \int\limits_{-\infty}^{\infty} \frac{d\theta'}{2 \pi i} \, G(\theta-\theta'+i\eta_-) \, L_-(\theta'-i \eta_-),
\end{split}
\end{equation}
where ${\cal M}$ denotes the fermion/soliton mass, $G(\theta)$ is the logarithmic derivative of the soliton-soliton scattering phase:
\begin{equation} \label{G}
G(\theta)=-i \, \frac{d}{d\theta} \log S_{++}^{++}(\theta)=\! \int\limits_{-\infty}^{\infty} \! d\omega \, 
e^{-i \, \omega \theta} \, \frac{\sinh(\tfrac{(p-1) \,\pi \omega}{2})}{2 \cosh( \tfrac{\pi \omega}{2}) 
\,\sinh(\tfrac{p \,  \pi \, \omega}{2})},
\end{equation}
$0<\eta_\pm<\text{min}(p \pi,\pi)$ are small positive contour deformation parameters, which must be smaller than 
the distance of the first pole of $G(\theta)$ from the real axis. For the case of nonzero chemical potential their minima cannot 
be arbitrarily small, as we will explain in the subsequent paragraphs.
The functions $L_\pm(\theta)$ denote the nonlinear combinations of $Z(\theta)$ as follows:
\begin{equation} \label{Lpm}
L_\pm(\theta)=\ln\left(1+ \, e^{\pm i \, Z(\theta)} \right).
\end{equation}
With the help of the solutions to (\ref{DDVtwist}), the energy and momentum of the ground state can be computed by the formulas:
\begin{equation} \label{Econt}
\begin{split}
E_0^{(\beta,\mu)}= -\frac{{\cal M}}{2 \pi i} \sum\limits_{\alpha=\pm} \, \alpha
\int\limits_{-\infty}^{\infty} d\theta \, \sinh(\theta+i \, \alpha\, \eta_\alpha) \, L_{\alpha}(\theta+i \, \alpha \, \eta_\alpha),
\end{split}
\end{equation}
\begin{equation} \label{Pcont}
\begin{split}
P_0^{(\beta,\mu)}= -\frac{{\cal M}}{2  \pi i} \sum\limits_{\alpha=\pm} \, \alpha
\int\limits_{-\infty}^{\infty} d\theta \, \cosh(\theta+i \, \alpha\, \eta_\alpha) \, L_{\alpha}(\theta+i \, \alpha \, \eta_\alpha).
\end{split}
\end{equation}
The equations (\ref{DDVtwist}), describe the continuum limit of a lattice fermion model with twisted boundary conditions:
\begin{equation}\label{lattPsiBC}
\begin{split}
\psi_{n+N}=-e^{-i \, \omega}\,  \psi_n, \qquad \bar{\psi}_{n+N}=-e^{i \, \omega\,} \bar{\psi}_n,
\end{split}
\end{equation}
where $N$ denotes the size of the discrete lattice.

To get the equations for the case of our interest, we should specify the value of $\omega$ in (\ref{DDVtwist}). This can be done by comparing 
the lattice (\ref{lattPsiBC}) and the continuum (\ref{twistedBC}) versions of the twisted boundary conditions.
For the correct comparison, we rephrase (\ref{lattPsiBC}) and (\ref{twistedBC}) such that the charge of the appropriate fields enter 
the boundary conditions and we translate (\ref{twistedBC}) into the space direction, because we applied the Euclidean trick:
\begin{equation}\label{QPsiBC}
\begin{split}
\psi_{n+N}&=-e^{i \, \omega\, Q_{\psi}^{(L)}} \psi_n, \qquad \bar{\psi}_{n+N}=-e^{i \, \omega\, Q_{\bar{\psi}}^{(L)}} \bar{\psi}_n, \\
\psi(x+\beta)&=-e^{-\beta \, \mu \, Q_{{\psi}}} \, \psi(x), \qquad 
\bar{\psi}(x+\beta)=-e^{-\beta \, \mu \, Q_{\bar{\psi}}} \, \bar{\psi}(x),
\end{split}
\end{equation}
where $Q_{{\psi}}^{(L)}=-1$ and $Q_{\bar{\psi}}^{(L)}=+1$ are the charges{\footnote{In the lattice, the fermion charge can be identified 
as twice the z-component of the spin operator: $Q^{(L)}=2 \,S_z.$ By the charge of a lattice fermion field we mean the eigenvalue under 
the commutator with the charge: $[Q^{(L)},\psi_n]=Q_\psi^{(L)} \, \psi_n,$ etc. }}
of the lattice fermion fields, and $Q_{{\psi}}=-1$ and $Q_{\bar{\psi}}=+1$ are those of the continuum fields. 

The identification of the continuum limit of the lattice phase factor with its continuum counterpart - each appearing in separate lines of equation (\ref{QPsiBC}) - leads to the following operator-valued equation:
\begin{equation}\label{omega1}
\begin{split}
i \omega \, Q^{(L)}_{cont}=-\mu \, \beta \, Q,
\end{split}
\end{equation}
where $Q^{(L)}_{cont}$ is the continuum limit of the bare lattice fermion charge operator, while $Q$ is that of the renormalized continuum theory. 
In \cite{Hegedus:2017muz}, it has been shown, that taking the continuum limit in the framework of the light-cone lattice approach, 
the two charges differ by a factor of $\frac{p+1}{p}.$ Namely, 
 \begin{equation}\label{QQL}
\begin{split}
Q^{(L)}_{cont}=\frac{p+1}{p}\, Q.
\end{split}
\end{equation}
Thus, (\ref{omega1}) and (\ref{QQL}) implies, that the twist constant in (\ref{DDVtwist}) takes the following form in the 
continuum theory:
\begin{equation}\label{omdet}
\begin{split}
\omega \, \frac{p+1}{p}=i\, \mu\, \beta.
\end{split}
\end{equation}
Thus, the NLIE for the case with chemical potential takes the form as follows:
\begin{equation} \label{DDVcont}
\begin{split}
Z(\theta)&={\cal M} \beta \, \sinh \theta +\,i \, \beta \, \mu +
\, \int\limits_{-\infty}^{\infty} \frac{d\theta'}{2 \pi i} \, G(\theta-\theta'-i\eta_+) \, L_+(\theta'+i \eta_+)- \\
&-\, \int\limits_{-\infty}^{\infty} \frac{d\theta'}{2 \pi i} \, G(\theta-\theta'+i\eta_-) \, L_-(\theta'-i \eta_-).
\end{split}
\end{equation}
Now, we are in the position to discuss the allowed range of the contour deformation parameters $\eta_\pm.$ 
It is well known from the literature of the NLIE \cite{Destri:1994bv}, that these parameters should be chosen to avoid 
those singularities of the functions $L_\pm(\theta),$ which lie along or very near to the real axis. 
When the chemical potential is zero, it is known, that the relevant singularities lie exactly on the real axis.  
Thus the only requirement for the minimum of $\eta_\pm,$ is that it should be larger than zero. On the other hand, when 
a chemical potential is also present in the theory, these singularities move away from the real axis and gain imaginary parts, as well. 
If $\mu>0,$ then the singularities move downwards to the lower half plane of the complex plane. In the opposite case they move to 
the upper half plane. This is very similar to the case of the XXZ-model \cite{Klumper:1993vq,Destri:1994bv,Essler2005}. 

Now, we give a more precise description to the allowed minimal value of the parameters $\eta_\pm.$  
Let $\theta_j$ the singularities of $L_\pm(\theta)$ in the strip: $-\text{min}(p\, \pi,\pi)< \text{Im} \, \theta< \text{min}(p\, \pi,\pi).$
They satisfy the Bethe equations:
\begin{equation}\label{BAcont}
\begin{split}
Z(\theta_j)=2 \pi \, I_j, \qquad \qquad I_j \in \mathbb{Z}+1/2.
\end{split}
\end{equation}
Then the following criterion can be given to the allowed range of the contour deformation parameters:
\begin{equation}\label{etas}
\begin{split}
0<\eta_+&<\text{min}(p\, \pi,\pi), \quad     \text{max}\, | \text{Im} \, \theta_j| <\eta_-<\text{min}(p\, \pi,\pi), \quad \text{if} \, \qquad \mu>0, \\
0<\eta_-&<\text{min}(p\, \pi,\pi), \quad     \text{max}\, | \text{Im} \, \theta_j| <\eta_+<\text{min}(p\, \pi,\pi), \quad \text{if} \, \qquad \mu<0. \\
\end{split}
\end{equation}
Though, these lower bounds cannot be a priori determined, one can assume, that at any values of the parameters $\beta$ and $\mu,$ 
it is possible escape from these singularities with the appropriate choice of $\eta_\pm.$ Which means, that introduction of extra source terms 
in the equations can always be avoided. 

To get a simple insight into the singularity structure in the neighborhood of the real axis, one can
 consider the equations at the free fermion point 
$p=1.$ Here, $G(\theta)=0,$  and the equation immediately gives the solution, as well:
\begin{equation}\label{Zp1}
\begin{split}
Z_{p=1}(\theta)&={\cal M} \beta \, \sinh \theta +\,i \, \beta \, \mu.
\end{split}
\end{equation}
The positions of the singularities from (\ref{BAcont}):
\begin{equation}\label{thj}
\begin{split}
\theta_j=\text{arcsinh} \left( \frac{2 \, \pi \, I_j}{{\cal M} \beta}-i \frac{\mu}{\cal M} \right), \qquad j\in \mathbb{Z}.
\end{split}
\end{equation}
So, it is easy to see, that the maximally deviating roots are $\theta_{\pm 1},$ and that 
the maximum of the deviation\footnote{Namely: $\eta_{\pm} >\underset{j }{\mbox{max}}|\theta_j|= 
\mbox{max}|\theta_{\pm 1}|\leq \underset{\mu}{\mbox{max}}|\theta_{\pm 1}|= \tfrac{\pi}{2}.$} 
from the real axis is $\frac{\pi}{2}.$ It happens, when $|\mu| \to \infty.$ This implies, that 
around the $p=1$ point, even in the worse case  
one can choose the values of $\eta_\pm$ to avoid the introduction of new source terms into the equations.

We close this section, with mentioning another property of $Z(\theta)$ deviating from the so far discussed cases in the literature. 
Namely, due to the twist parameter in (\ref{DDVcont}) being imaginary, $Z(\theta)$ will not stay to be a real analytic 
function, which means, that $Z(\theta)^*\neq Z(\theta^*),$ where $^*$ stands for complex conjugation. Again this fact is 
completely analogous to the NLIE description to the  thermodynamics of the XXZ-model \cite{Klumper:1993vq,Destri:1994bv,Essler2005}. 
Actually, if the dependence on the chemical potential is made explicit, the conjugation properties:   $Z(\theta|\mu)^*=Z(\theta^*|-\mu)=-Z(-\theta^*|\mu)$  hold for $\mu \in \mathbb{R}$.

\section{A few simple tests} \label{sect4}

In this section we perform a few simple tests on the NLIE (\ref{DDVcont}).

\subsection{The UV limit}

In the ultraviolet limit, when $\beta \to 0,$ two simple cases can be considered. 
 In both cases the 
equations can be treated analytically in the UV in the usual way using the plateau argument 
and the dilogarithm trick \cite{Destri:1994bv}. 

First, we consider the $\beta \to 0,$ while $\mu=$ finite,  limit. In this limit,  
the ground state energy becomes that of a $c=1$ conformal field theory \cite{Destri:1994bv,Feverati:2000xa}:
\begin{equation}\label{Econfc1}
\begin{split}
E_0^{(\beta,\mu)} \to -\frac{\pi}{6 \beta}.
\end{split}
\end{equation}
For a simple, numerical check, see table \ref{text1}.
in appendix \ref{appA}. 

On the other hand one can consider the limit, when $\mu \beta=\mu_0$ is kept finite, when $\beta \to 0.$ 
In this case, the UV formula for the twisted case of  \cite{Destri:1994bv,Feverati:2000xa} can be used, and one obtains:
\begin{equation}\label{Econfmu0}
\begin{split}
E_0^{(\beta,\mu)} \to -\frac{\pi}{6 \beta} \, c_{eff}(\mu_0), \qquad \qquad c_{eff}(\mu_0)=1+\frac{6 \, p}{p+1} \, \frac{\mu_0^2}{\pi^2}.
\end{split}
\end{equation}
By the numerical solution to the equations, it is easy to check this limiting value, too. 
Some confirming numerical data, can be found in table \ref{text2}.  
in appendix \ref{appA}. 

\subsection{An IR computation}

In this subsection, we perform a very simple calculation in the infrared limit, when $\beta \to \infty,$ and $\mu \, \beta \to \mu_0=\text{finite}.$
The purpose of this short subsection is to demonstrate, that the NLIE gives identical result to the TBA formulation of the model \cite{Nagy:2023phz}. 
The actual form of the TBA equations depend strongly on the value of the coupling constant. Thus, at certain values they can have 
a lot of components and a very complicated form. To simplify the comparison, we consider the problem in the  $p=1/2$ point, which is 
the simplest special point, where the  soliton-antisoliton scattering becomes reflectionless.
We do so, because in the reflectionless points with $p=1/(n_B+2)$, the TBA equations for the model are simple, and can be formulated as follows \cite{Zamolodchikov:1991et,Nagy:2023phz}:
\begin{equation}\label{TBAeqs}
\begin{split}
\epsilon_a(\theta)&=\nu_a(\theta)-\sum\limits_{b=1}^{n_B+2} \int\limits_{-\infty}^{\infty} d\theta' \, \varphi_{ab}(\theta-\theta') 
\, \log(1+e^{-\epsilon_b(\theta')}), \qquad a=1,..,n_B+2,   \\
\nu_a(\theta)&=m_a \, \beta \, \cosh(\theta)-q_a \, \mu \, \beta, \\
E^{(\beta,\mu)}_0&=-\frac{1}{2 \pi} \, \sum\limits_{a=1}^{n_B+2} \, m_a  \int\limits_{-\infty}^{\infty} d\theta \, \cosh(\theta) \, 
\log(1+e^{-\epsilon_a(\theta)}), \\
P^{(\beta,\mu)}_0&=-\frac{1}{2 \pi} \, \sum\limits_{a=1}^{n_B+2} \, m_a  \int\limits_{-\infty}^{\infty} d\theta \, \sinh(\theta) \, 
\log(1+e^{-\epsilon_a(\theta)}).
\end{split}
\end{equation}
where $n_B$ denotes the number of breathers entering the mass spectrum. 
\begin{equation}\label{qa}
\begin{split}
q_0=0, \qquad a=1,..,n_B, \qquad q_{n_B+1}=-q_{n_B+2}=1,
\end{split}
\end{equation}
are the fermion/topological charges corresponding to the breathers and to the soliton and  
antisoliton{\footnote{Here, we used the sine-Gordon terminology, and call the fermion and antifermion as soliton and antisoliton}} respectively.
The mass parameters correspond to the masses of the particles in the theory:
\begin{equation}\label{ma}
\begin{split}
m_a=m_{B_a}, \qquad a=1,...,n_B, \qquad m_{n_B+1}=m_{n_B+2}={\cal M},
\end{split}
\end{equation}
with $m_{B_a}$ corresponding to the mass of the $a$th breather:
\begin{equation}\label{mBa}
\begin{split}
m_{B_a}=2 \, {\cal M} \, \sin\frac{\pi \, a \,p}{2}, \qquad a=1,..,n_B.
\end{split}
\end{equation}
Similarly to the mass and charge parameters of the theory, the pseudoenergies $\epsilon_a(\theta)$ in the (\ref{TBAeqs}) 
correspond to the $a$th particle in the spectrum, and the kernel matrix $\varphi_{ab}(\theta)$ is defined form the scattering matrix 
of particle species $a$ and $b,$ by the formula:
\begin{equation}\label{fiab}
\begin{split}
\varphi_{ab}(\theta)=\frac{1}{2 \pi i} \, \frac{d}{d \theta} \, \log S_{ab}(\theta),
\end{split}
\end{equation}
where the necessary S-matrix elements are given below. The elementary building block is the function:
\begin{equation}\label{S_a}
\begin{split}
S_a(\theta)=\frac{\sinh \theta+ i \, \sin(\pi a)}{\sinh \theta- i \, \sin(\pi a).}
\end{split}
\end{equation}
Then the necessary matrix elements can be listed. The soliton-soliton, soliton-breather, and breather-breather matrix 
elements takes the form in the reflectionless points, as follows \cite{Zamolodchikov:1978xm}:
\begin{equation}\label{Selemets}
\begin{split}
S_{\pm,\pm}(\theta)&=(-1)^{\frac{1}{p}} \, \prod_{k=1}^{1/p-1} \frac{\cosh\left(\frac{\theta-i \, \pi \, k \, p}{2} \right)}
{\cosh\left(\frac{\theta+i \, \pi \, k \, p}{2} \right)},  \\
S_{\pm,B_k}&=\prod\limits_{j=0}^{k-1} \, S_{\frac{1-k \, p}{2}+p \, j}(\theta), \\
S_{B_k,B_{k'}}&=\frac{\prod\limits_{j=0}^{k'} \, S^2_{\frac{k-k'}{2}\, p+j \, p}(\theta) }{ S_{\frac{k-k'}{2}\, p}(\theta) \, S_{\frac{k+k'}{2}\, p}(\theta) }, \qquad 
k\geq k'.
\end{split}
\end{equation}
Now, we can continue with the description of the computation in the IR limit upto the order of $e^{-2 {\cal M} \beta}$ at the $p=1/2$ value of the coupling constant.
From now on, on the one hand, to get rid of writing out unnecessarily a lot of ${\cal M}$ factors, we do the computations in the ${\cal M}=1$ unit. 
On the other hand, we scale $\mu$ such, that  $\mu_0=\mu \, \beta$ be a $\beta$ independent finite number. Then, 
we expand the equations in the $\beta \to \infty$ limit. 

At this special point the kernel takes a particularly simple form: 
\begin{equation}\label{G2}
\begin{split}
G_2(\theta)=\frac{1}{2 \pi } G(\theta)_{p=1/2}=-\frac{1}{ 2 \, \pi \, \cosh(\theta)}. \qquad
\end{split}
\end{equation}
The important property of this function, which we will use is that it has simple poles at $\pm \, i \, \tfrac{\pi}{2},$ in the strip 
$-\pi<\text{Im} \theta<\pi,$ with residues:
\begin{equation}\label{resG2}
\begin{split}
\text{Res}_{\theta=i \tfrac{i\, \pi}{2}} \, G_2(\theta)=\frac{i}{2 \,  \pi}, \qquad \text{Res}_{\theta=-i \tfrac{i\, \pi}{2}} \, G_2(\theta)=-\frac{i}{2\, \pi} .
\end{split}
\end{equation}
In the IR limit the functions $e^{\pm i \, Z(x\pm \, i \, \eta)}$ become exponentially small in $\beta.$ So, we will expand in them upto second order. 
In this limit, the solution to the NLIE can be written as follows\footnote{In what follows, unless stated 
otherwise all integrals without specified bounds are assumed to range from $-\infty$ to $\infty$.}:
\begin{equation}\label{Zser1}
\begin{split}
Z(\theta)&=Z_0(\theta)+\delta Z(\theta)+..., \\
Z_0(\theta)&=\beta \, \sinh \theta +i \, \mu_0, \\
\delta Z(\theta)&=\frac{1}{i} \int dx \, G_2(\theta-x-i \, \eta')\, e^{i \, Z_0(x+i \, \eta')}-\frac{1}{i} \int dx \, G_2(\theta-x+i \, \eta')\, e^{-i \, Z_0(x-i \, \eta')}.
\end{split}
\end{equation}
Expanding the $\log$ in the energy (\ref{Econt}), one ends up with 3 relevant contributions:
\begin{equation}\label{E01}
\begin{split}
E_0^{(\beta,\mu)}&=E_0^{(1)}+E_0^{(2)}+\delta E_0^{\delta Z}+...,\\
E_0^{(1)}&=\frac{-1}{2 \, \pi \, i} 
\sum\limits_{\alpha=\pm} \alpha \! \int dx  \, \sinh(x\!+i \, \alpha \, \eta) \,  e^{i\, \alpha \, Z_0(x+i  \,\alpha\,  \eta)}, \\
E_0^{(2)}&=\frac{1}{4 \, \pi \, i} 
\sum\limits_{\alpha=\pm} \alpha \! \int dx  \, \sinh(x\!+i \, \alpha \, \eta) \,  e^{2 \,i\, \alpha \, Z_0(x+i  \,\alpha\,  \eta)}, \\
\delta E_0^{\delta Z}&=\frac{-1}{2 \, \pi }   \sum\limits_{\alpha=\pm}  
 \int dx  \sinh(x\!+i \, \alpha \, \eta) \,  e^{i \, \alpha  Z_0(x+i  \, \alpha \, \eta)} \, \delta Z(x+i \, \alpha 
\, \eta).
\end{split}
\end{equation}
In the first two terms, one can  shift the contours by taking  $\eta \to \frac{\pi}{2},$ without meeting any 
singularities.  As a result, 
one ends up with the following formulas being compatible with the IR expansion coming from the TBA:
\begin{equation}\label{E1E2}
\begin{split}
E_0^{(1)}&=\frac{-1}{2 \pi} \, (e^{\mu_0}+e^{-\mu_0}) \, \int dx \, \cosh x\, e^{-\beta  \cosh x,}, \\
E_0^{(2)}&=\frac{1}{4 \pi} \, (e^{2 \mu_0}+e^{-2 \, \mu_0}) \, \int dx \, \cosh x\, e^{-2 \,\beta \cosh x,}.
\end{split}
\end{equation}
These terms correspond to the one and two, pure soliton and antisoliton contributions, which doesnot contain the  S-matrix data. A factor $e^{\mu_0}$ comes from the 
soliton and a factor $e^{-\mu_0}$ comes from the antisoliton contributions, respectively.  
The S-matrix dependence is encoded into the form of the terms containing $\delta Z(\theta).$
\begin{equation}\label{EdZ}
\begin{split}
\delta E_0^{\delta Z}=\delta E_{0+}^{\delta Z}+\delta E_{0-}^{\delta Z}, \qquad \qquad  \\
\text{with:} \qquad \qquad \delta E_{0+}^{\delta Z}&=\delta E_{0++}^{\delta Z}+\delta E_{0+-}^{\delta Z}, \\
\delta E_{0-}^{\delta Z}&=\delta E_{0-+}^{\delta Z}+\delta E_{0--}^{\delta Z},
\end{split}
\end{equation}
where
\begin{equation}\label{dE+-}
\begin{split}
\delta E_{0++}^{\delta Z}=\frac{-1}{2 \pi i} \int_{\Gamma_{\eta}} dz \int_{\Gamma_{\eta'}} dz' \, \sinh z \, e^{i \,\beta \, \sinh z -\mu_0} \, 
\, e^{i \, \beta \, \sinh z' -\mu_0} \, G_2(z-z'), \\
\delta E_{0+-}^{\delta Z}=\frac{1}{2 \pi i} \int_{\Gamma_{\eta}} dz \int_{\Gamma_{-\eta'}} dz' \, \sinh z \, e^{i \, \beta \, \sinh z } \, 
\, e^{-i \, \beta \, \sinh z' } \, G_2(z-z'), \\
\delta E_{0-+}^{\delta Z}=\frac{-1}{2 \pi i} \int_{\Gamma_{-\eta}} dz \int_{\Gamma_{\eta'}} dz' \, \sinh z \, e^{-i \,\beta \, \sinh z } \, 
\, e^{i \, \beta \, \sinh z' } \, G_2(z-z'), \\
\delta E_{0--}^{\delta Z}=\frac{1}{2 \pi i} \int_{\Gamma_{-\eta}} dz \int_{\Gamma_{-\eta'}} dz' \, \sinh z \, e^{-i \, \beta \, \sinh z +\mu_0} \, 
\, e^{-i \, \beta \, \sinh z' +\mu_0} \, G_2(z-z'), \\
\end{split}
\end{equation}
 where $\Gamma_\eta$ is a straight contour with distance $\eta$ from the real axis of the complex plane{\footnote{Namely, $\Gamma_{\eta}(x)=x+i \eta,$ $x \in \mathbb{R}.$ }}.
 The contour deformation parameters are positive and satisfy the inequality:
 $0<\eta_{min}<\eta'<\eta<\tfrac{\pi}{2},$ where $\eta_{min}$ denotes the maximal deviation of the singularities of $L_\pm(\theta)$ from the real axis. 
In $\delta E_{0++}^{\delta Z}$ and $\delta E_{0--}^{\delta Z}$ the two integration contours run in the same half-plane. First, one deforms the 
$\Gamma_{\pm \eta}$ contours to $\Gamma_{\pm \tfrac{\pi}{2}}$ by a shift $\eta \to \tfrac{\pi}{2}.$ It is easy to see, that no singularity is enclosed by these shifts.  
Then, one deforms the $\Gamma_{\pm \eta'}$ contours to $\Gamma_{\pm \tfrac{\pi}{2}}$ by a shift $\eta' \to \tfrac{\pi}{2}.$ Again no singularities are enclosed, giving the results:
\begin{equation}\label{dE++--}
\begin{split}
\delta E_{0++}^{\delta Z}=\frac{-1}{2 \pi} \int dx \, \cosh(x) \int dx' \, G_2(x-x') \, \, e^{-\beta \, \cosh (x) -\beta \, \cosh (x')-2\mu_0}, \\
\delta E_{0--}^{\delta Z}=\frac{-1}{2 \pi} \int dx \, \cosh(x) \int dx' \, G_2(x-x') \, \, e^{-\beta \, \cosh (x) -\beta \, \cosh (x')+2\mu_0}.
\end{split}
\end{equation}
These are the pure two soliton and two antisoliton contributions, which contain the S-matrix data. 
The $\delta E_{0+-}^{\delta Z}$ and $\delta E_{0-+}^{\delta Z}$ terms contain the chargeless 
contributions, coming from the breather, and the soliton-antisoliton states.
First, consider $\delta E_{0+-}^{\delta Z}$. When pushing the contour $\Gamma_{\eta}$ towards $\Gamma_{\tfrac{\pi}{2}},$ a pole coming from the kernel 
at $z=z'+i \tfrac{\pi}{2}$ is encircled, giving the result:
\begin{equation}\label{dE0+-v1}
\begin{split}
\delta E_{0+-}^{\delta Z}&=\frac{-1}{2 \pi} \int dx  \cosh(x) \, e^{-\beta \, \cosh(x)} \, 
\int_{\Gamma_{-\eta'}} dz'  \, e^{-i \, \beta \, \sinh z' }  \, G_2(x+i \, \tfrac{\pi}{2}-z')-\\
&-\frac{1}{2 \pi \, i} \int_{\Gamma_{-\eta'}} dz' \sinh(z'+i \, \tfrac{\pi}{2}) \, e^{i \, \beta \, \sinh(z'+i \tfrac{\pi}{2})-i \, \beta \, \sinh (z')}.
\end{split}
\end{equation}
Finally, in both integrals the contour in $z'$ is pushed to saddle point gives the final result for $\delta E_{0+-}^{\delta Z}.$ A very similar deformation 
of the contours lead to a similar result for $\delta E_{0-+}^{\delta Z}.$ The final formula for them is as follows:
\begin{equation}\label{dE0+--+}
\begin{split}
\delta E_{0+-}^{\delta Z}&=\frac{-1}{2 \pi i} \int dx \, \sinh(x+i \tfrac{\pi}{4}) \, e^{-2 \beta \, \sin(\tfrac{\pi}{4}) \, \cosh x}+ \\
&+\frac{1}{2 \pi} \int dx \cosh x \, \int dx' \, G_2(x-x'+i \, \pi)\, e^{-\beta \cosh(x)-\beta \cosh(x')} \\
\delta E_{0-+}^{\delta Z}&=\frac{1}{2 \pi i} \int dx \, \sinh(x-i \tfrac{\pi}{4}) \, e^{-2 \beta \, \sin(\tfrac{\pi}{4}) \, \cosh x}+ \\
&+\frac{1}{2 \pi} \int dx \cosh x \, \int dx' \, G_2(x-x'-i \, \pi)\, e^{-\beta \cosh(x)-\beta \cosh(x')} \\
\end{split}
\end{equation}
Adding all contributions together, one ends up with the final result for $\delta E_0^{\delta Z}:$
\begin{equation}\label{dE0dZfin}
\begin{split}
\delta E_0^{\delta Z}&=\frac{-1}{2 \pi} \int dx \int dx' \, \cosh(x) \, G_2(x-x') \, e^{-\beta \, \cosh(x)-\beta \, \cosh(x') -2 \mu_0}-\\
&-\frac{1}{2 \pi} \int dx \int dx' \, \cosh(x) \, G_2(x-x') \, e^{-\beta \, \cosh(x)-\beta \, \cosh(x') +2 \mu_0}+\\
&+\frac{1}{2 \pi} \int dx \int dx' \, \cosh(x) \, \left[ G_2(x-x'-i \pi)+G_2(x-x'+i \pi) \right] \, e^{-\beta \, \cosh(x)-\beta \, \cosh(x') } \\
&-\frac{m_{B_1}}{2 \pi} \int dx  \, \cosh(x) \,  \, e^{-m_{B_1} \beta \, \cosh(x)},
\end{split}
\end{equation}
where we exploited, that at $p=1/2,$ the breather mass takes the form: $m_{B_1}=2 {\cal M} \sinh(\tfrac{\pi}{4}).$
The sum of  (\ref{E1E2}) and (\ref{dE0dZfin}) gives the result for the large $\beta$ expansion of the energy derived from the NLIE, upto $e^{-2 {\cal M} \beta }$ order. 
Iterating the TBA equations (\ref{TBAeqs}) at large $\beta$, it is straightforward to derive this result. 

Though, this IR computation was carried out at a very specific value of the coupling constant, it is still enlightening enough, because 
beyond showing the agreement between the TBA and NLIE results, it reveals, how the breather contributions arise in the framework of the NLIE.
Namely, typically they arise from the product of two exponentials of $Z(x):$   where one is lying 
in an upper half plane contour, and the other one, on a lower half plane contour, $\sim e^{i \, Z(x+i \, \eta)}\, e^{-i \, Z(x'-i \, \eta')}$. This combination 
ensures the zero charge (zero chemical potential) of the expression, and the breather contributions 
come from the residue of the poles of the kernel $G(\theta).$

\section{Expectation values} \label{sect5}

Having the free-energy at hand, it is straightforward to compute, the expectation values 
of the conserved charges. At least only those, which are contained in the definition of the free energy.  
In an integrable model in principle all conserved charges can be brought into the definition of the free-energy, 
giving the so-called Generalized Gibbs Ensemble (GGE). In this paper, we can't extend the NLIE formalism to 
describe the GGE, but we can add one more term to the NLIE to account for the momentum 
conservation\footnote{We attempted to naively add source terms proportional to $\sim \sinh(3 \theta)$ or $\sim \cosh(3 \theta)$ to the equations in order to account for the next higher-spin charges. However, these attempts failed when compared to the results of the TBA. We believe that, if such an extension is possible, then even the source terms must exhibit more intricate analytic properties than what such a naive guesswork would suggest.}. 
In this case the NLIE takes the form:
\begin{equation} \label{DDVcont1}
\begin{split}
Z(\theta)&={\cal M} \beta \, \sinh \theta +\lambda {\cal M} \beta \, \cosh \theta +\,i \, \beta \, \mu+ \\
&+\, \int\limits_{-\infty}^{\infty} \frac{d\theta'}{2 \pi i} \, G(\theta-\theta'-i\eta_+) \, L_+(\theta'+i \eta_+) 
-\, \int\limits_{-\infty}^{\infty} \frac{d\theta'}{2 \pi i} \, G(\theta-\theta'+i\eta_-) \, L_-(\theta'-i \eta_-),
\end{split}
\end{equation}
where $\lambda$ is the coefficient of the momentum term. The analogous modification of the TBA equations (\ref{TBAeqs}),
is a simple modification of the source function by a momentum density term:
\begin{equation}\label{nuamod}
\begin{split}
\nu_a(\theta)&=m_a \, \beta \, \cosh(\theta)+m_a \, \lambda \, \beta \, \sinh(\theta)-q_a \, \mu \, \beta.
\end{split}
\end{equation}
The NLIE converges, if $-1<\lambda<1.$ In the reflectionless points, where the TBA has the simple form (\ref{TBAeqs}), 
it is easy to check numerically, that the TBA with (\ref{nuamod}) and the NLIE (\ref{DDVcont1}) give identical results. 
Though, the form of both the TBA and the NLIE have been changed, the formulas for the energy (\ref{Econt}), 
and momentum (\ref{Pcont}), (\ref{TBAeqs}) remain the same, but now become $\lambda$ dependent. 

 The expectation values of the conserved currents can be straightforwardly computed following the general 
rules of thermodynamics. The energy (\ref{Econt}) and the momentum (\ref{Pcont}) of the mirror rotated 
theory, will serve as the generators of the expectation values of the conserved currents, in the following way. 
First one introduces the thermodynamic conjugate variables: $\beta^{(e)}=\beta,$ 
$\beta^{(p)}=\lambda \, \beta, $ $\beta^{(\mu)}=\mu \, \beta, $ associated with the energy, momentum, and 
topological charge, respectively, and and reparametrizes the relevant equations accordingly.  
The resulting NLIE takes the form: 
\begin{equation} \label{DDVTconj}
\begin{split}
Z(\theta)&=S(\theta|\beta^{(e)},\beta^{(p)},\beta^{(\mu)}) + \\
&+\, \int\limits_{-\infty}^{\infty} \frac{d\theta'}{2 \pi i} \, G(\theta-\theta'-i\eta_+) \, L_+(\theta'+i \eta_+) 
-\, \int\limits_{-\infty}^{\infty} \frac{d\theta'}{2 \pi i} \, G(\theta-\theta'+i\eta_-) \, L_-(\theta'-i \eta_-),\\
&\text{with: } 
S(\theta|\beta^{(e)},\beta^{(p)},\beta^{(\mu)})={\cal M} \, \beta^{(e)} \, \sinh(\theta)+{\cal M} \,  \beta^{(p)} \, \cosh(\theta)+i \, \beta^{(\mu)}.
\end{split}
\end{equation}
Similarly, in the TBA framework, the source term is modified as:
\begin{equation}\label{nuamodT}
\begin{split}
\nu_a(\theta|\beta^{(e)},\beta^{(p)},\beta^{(\mu)})&=m_a \, \beta^{(e)} \, \cosh(\theta)+m_a \,  \beta^{(p)} \, \sinh(\theta)-q_a \, \beta^{(\mu)}.
\end{split}
\end{equation}
Since the ground state energy of the mirror rotated model, is nothing but the ratio of free-energy density to  
temperature in the original model, it will serve as the generator of the expectation values of the charge densities  or 
equivalently the time oriented components of the currents. Namely, the expectation value of the charge $h:$ 
can be computed from the energy as follows \cite{Nagy:2023phz}:
\begin{equation}\label{hexp}
\begin{split}
\langle h \rangle_\beta=\partial_{\beta^{(h)}} E_0(\beta^{(e)},\beta^{(p)},\beta^{(\mu)}), \qquad h \in \{e,p,\mu\}.
\end{split}
\end{equation}
Similarly the momentum of the mirror rotated theory, corresponds to the free-energy flux \cite{Castro-Alvaredo:2016cdj,Doyon:2019nhl}, 
which generates the expectation values of the currents, i.e. the space oriented components of the 
conserved currents:
\begin{equation}\label{jhexp}
\begin{split}
\langle j_h \rangle_\beta=\partial_{\beta^{(h)}} P_0(\beta^{(e)},\beta^{(p)},\beta^{(\mu)}), \qquad h \in \{e,p,\mu\}.
\end{split}
\end{equation}
These, formulas lead to expressions in terms of solutions of certain linear integral equations for these 
expectation values. Before turning to them, it is worth mentioning a few trivial statements. 
When $\lambda=\beta^{(p)}=0,$ then the expectation value of the current, associated to the 
topological/fermion charge ($\langle j_1 \rangle_\beta$) becomes zero. This is because in the mirror 
rotated theory it corresponds to the finite volume vacuum expectation value of the charge density. 
But, the vacuum is chargeless, so this expectation value must be zero. 

Similarly in the $\lambda=\beta^{(p)}=0$ case, the momentum of the vacuum of the mirror rotated theory 
becomes zero, which implies, that the expectation values of the currents associated to the energy and the 
topological charge 
become zero.

Now, as an example  we will discuss the formulas for the expectation values of the 
 charges and the corresponding currents.  

Using (\ref{hexp}), (\ref{jhexp}) and (\ref{Econt}), (\ref{Pcont}), the  
following formulas arise for the expectation values:
\begin{equation}\label{hexpZ}
\begin{split}
\langle h \rangle_\beta=-\frac{{\cal M}}{2 \, \pi} \sum\limits_{\alpha=\pm} \int\limits_{-\infty}^\infty 
d\theta \sinh(\theta+i \, \alpha \,  \eta_\alpha) \,  Z_h'(\theta+i \, \alpha \, \eta_\alpha) 
{\cal F}_\alpha(\theta+i \, \alpha \, \eta_\alpha), \\
\langle j_h \rangle_\beta=-\frac{{\cal M}}{2 \, \pi} \sum\limits_{\alpha=\pm} \int\limits_{-\infty}^\infty 
d\theta \cosh(\theta+i \, \alpha \,  \eta_\alpha) \,  Z_h'(\theta+i \, \alpha \, \eta_\alpha) 
{\cal F}_\alpha(\theta+i \, \alpha \, \eta_\alpha),
\end{split}
\end{equation}
where we introduced the notations:
\begin{equation}\label{calF}
\begin{split}
Z_h'(\theta)=\partial_{\beta^{(h)}} Z(\theta), \qquad \qquad 
{\cal F}_\pm(\theta)=\frac{e^{\pm i\, Z(\theta)}}{1+e^{\pm i\, Z(\theta)}}.
\end{split}
\end{equation}
Following from (\ref{DDVTconj}), the functions $Z_h'(\theta),$ satisfy the linear integral equations: 
\begin{equation}\label{Zvhlin}
\begin{split}
Z_h'(\theta)&=s_h(\theta)+ \sum\limits_{\alpha=\pm} 
\int\limits_{-\infty}^\infty \frac{dx}{2 \pi} G(\theta-x-i \alpha \, \eta_\alpha) \, {\cal F}_\alpha (x+i \, \alpha \, \eta_\alpha) 
\, Z_h'(x+i \, \alpha \, \eta_\alpha),  \\
s_h(\theta)&=\partial_{\beta^{(h)}} S(\theta|\beta^{(e)},\beta^{(p)},\beta^{(\mu)}), \qquad h \in \{e,p,\mu \}.
\end{split}
\end{equation}
Concretely,
\begin{equation}\label{sepmu}
\begin{split}
s_e(\theta)={\cal M} \, \sinh(\theta), \qquad s_p(\theta)={\cal M} \, \cosh(\theta), \qquad s_\mu(\theta)=i.
\end{split}
\end{equation}
Now, we apply these formulas to the case of topological charge and current to compare them to 
the earlier formulas in \cite{Hegedus:2017muz} for the finite volume expectation values of the $U(1)$ current. 

In the thermodynamical formalism of this paper, from (\ref{hexpZ}) 
the expectation values take the form as follows:
\begin{equation}\label{JexpZ}
\begin{split}
\langle j_0 \rangle_\beta=-\sum\limits_{\alpha=\pm} \int\limits_{-\infty}^\infty 
\! \frac{d\theta}{2 \pi} \, s_e(\theta+i \, \alpha \,  \eta_\alpha) \,  Z_\mu'(\theta+i \, \alpha \, \eta_\alpha) 
{\cal F}_\alpha(\theta+i \, \alpha \, \eta_\alpha), \\
\langle j_1 \rangle_\beta=- \sum\limits_{\alpha=\pm} \int\limits_{-\infty}^\infty 
\! \frac{d\theta}{2 \pi} \,  s_p(\theta+i \, \alpha \,  \eta_\alpha) \,  Z_\mu'(\theta+i \, \alpha \, \eta_\alpha) 
{\cal F}_\alpha(\theta+i \, \alpha \, \eta_\alpha).
\end{split}
\end{equation}
Using the identity \cite{Hegedus:2017muz}:
\begin{equation}\label{identity}
\begin{split}
&\sum\limits_{\alpha=\pm} \int\limits_{-\infty}^\infty \frac{d \theta}{2 \pi} s_h(\theta+i \, \alpha \, \eta_\alpha) 
Z_{h'}'(\theta+i \, \alpha \, \eta_\alpha) \, {\cal F}_\alpha(\theta+i \, \alpha \, \eta_\alpha)=\\
&\sum\limits_{\alpha=\pm} \int\limits_{-\infty}^\infty \frac{d \theta}{2 \pi} s_{h'}(\theta+i \, \alpha \, \eta_\alpha) 
Z_{h}'(\theta+i \, \alpha \, \eta_\alpha) \, {\cal F}_\alpha(\theta+i \, \alpha \, \eta_\alpha),
\qquad   h, h' \in \{e,p,\mu\}.
\end{split}
\end{equation}
The formula (\ref{JexpZ}), can be rephrased in a way reminiscent to those obtained in \cite{Hegedus:2017muz} for the 
expectation values of the $U(1)$ current in a finite box{\footnote{Taking into account the $j_0 \leftrightarrow j_1$ exchange of the currents, under the mirror 
rotation,too.}}:
\begin{equation}\label{JexpZha}
\begin{split}
\langle j_0 \rangle_\beta=- i \, \sum\limits_{\alpha=\pm} \int\limits_{-\infty}^\infty 
\! \frac{d\theta}{2 \, \pi} \,  \,  Z_e'(\theta+i \, \alpha \, \eta_\alpha) 
{\cal F}_\alpha(\theta+i \, \alpha \, \eta_\alpha), \\
\langle j_1 \rangle_\beta=-i\, \sum\limits_{\alpha=\pm} \int\limits_{-\infty}^\infty 
\! \frac{d\theta}{2 \, \pi} \,  Z_p'(\theta+i \, \alpha \, \eta_\alpha) 
{\cal F}_\alpha(\theta+i \, \alpha \, \eta_\alpha).
\end{split}
\end{equation}
The formula (\ref{JexpZha}) is appropriate to show, that when $\lambda=\beta^{(p)}=0,$ then 
$\langle j_1 \rangle_\beta=0.$ This is because in this case $Z_p'(\theta) \sim Z'(\theta).$ 
Thus, the expectation value becomes the integral of a total derivative of a function, which decays 
exponentially rapidly to zero at the infinities. 

\subsection{Expectation values of local operators}

Using the techniques described  in \cite{Jimbo:2010jv} and \cite{Hegedus:2019rju}, 
the NLIE formalism allows one not only to compute the 
expectation values of the currents, but 
to compute the expectation values of local operators, as well. 
The approach of \cite{Jimbo:2010jv} uses the fermionic basis for the local 
operators \cite{Boos:2006mq,Boos:2008rh,Jimbo:2008kn,Boos:2010qii}. 
The  formulas given in \cite{Jimbo:2010jv} and \cite{Hegedus:2019rju} can be straightforwardly 
generalized to our case with chemical potential. Only, the NLIE describing 
the sandwiching state should be changed to (\ref{DDVcont1}) and in contrary to the 
zero chemical potential case, the contours cannot be pushed arbitrarily close to the 
real axis. (I.e. $\eta_\pm>\eta_{min}>0.$)  
 For the computation of the expectation value of an arbitrary local operator, the interested reader 
should consult the papers \cite{Jimbo:2010jv} and \cite{Hegedus:2019rju}. 
In this paper, we restrict our attention to the expectation values of the vertex operators 
$e^{i \, m \, \beta_{SG} \,\Phi(x)}$ in the sine-Gordon model, with $m \in \mathbb{N},$ 
since they correspond to  experimentally measurable coherence factors in tunnel-coupled cold atomic condensates \cite{Hofferberth:2007qfs,Esslernincs}. To keep the formulas as short as possible, we introduce the notation:
$$ \nu=\frac{1}{p+1}.$$
The expectation values of the fields $e^{i \, m \, \beta_{SG} \,\Phi(x)},$ are given in the following way  
 \cite{Jimbo:2010jv,Hegedus:2019rju}: 
\begin{equation}  \label{primVEV}
\begin{split}
\langle  e^{i \, m \, \beta_{SG} \,\Phi(x)} \rangle=
{\kappa}^{-2 \, m^2\,(1-\tfrac{1}{\nu})}\, C_m(0) \, \underset{1 \leq j,k \leq m }{\mbox{det}} \, \Omega_{kj}, \qquad m=1,2,...
\end{split}
\end{equation}
where $C_m(\alpha)$ is a constant given by the formula for $m>0,$ as follows:
\begin{equation} \label{Cm}
\begin{split}
C_m(\alpha)=\prod\limits_{j=0}^{m-1} C_1(\alpha+2j\tfrac{1-\nu}{\nu}),
\end{split}
\end{equation}
with
\begin{equation} \label{C1}
\begin{split}
&C_1(\alpha)=i \, \nu \Gamma(\nu)^{4 x(\alpha)}\, \frac{\Gamma(-2 \nu x(\alpha))}{\Gamma(2 \nu x(\alpha))}\, \frac{\Gamma(x(\alpha))}{\Gamma(x(\alpha)+1/2)} \,
\frac{\Gamma(-x(\alpha)+1/2)}{\Gamma(-x(\alpha))} \cot(\pi x(\alpha)), \\
&\text{and} \qquad x(\alpha)=\frac{\alpha}{2}+\frac{1-\nu}{2 \nu}.
\end{split}
\end{equation}
The matrix elements of $\Omega_{j,k}$ are expressed in terms of a more fundamental 
matrix $\om_{2k-1,1-2j},$  as follows: 
\begin{equation}  \label{OM}
\begin{split}
\Omega_{j,k}=\om_{2k-1,1-2j}+\frac{i}{\nu} \, \delta_{j,k}\, 
\cot \left[ \frac{\pi}{2 \nu} (2k-1)\right], \qquad j,k=1,...m.
\end{split}
\end{equation}
The parameter $\kappa$ in (\ref{primVEV}), is nothing, but the coupling constant in the action (\ref{sG_Lagrangian}) of the sine-Gordon model. It is related to the soliton mass ${\cal M},$ by the formula \cite{Zamolodchikov:1995xk}:
\begin{equation} \label{muM}
\begin{split}
\kappa={\cal M}^\nu \, \Pi(\nu)^\nu, \qquad \text{where} \qquad 
\Pi(\nu)=\frac{\sqrt{\pi}}{2} \frac{\Gamma\left(\tfrac{1}{2 \nu}\right)}{\Gamma\left(\tfrac{1-\nu}{2 \nu}\right)} \, \Gamma\left( \nu \right)^{-\tfrac{1}{\nu}}.
\end{split}
\end{equation}

The only remaining undefined quantity in (\ref{primVEV}) is $\om_{2k-1,1-2j}.$
 It is expressed as an integral involving certain functions, 
which themselves are solutions to linear integral equations. These, in turn, involve  
the solution of the NLIE (\ref{DDVcont}) as part of the integration measure. Namely,
\begin{equation}  \label{omtom}
\begin{split}
\om_{2k-1,1-2j}=\frac{i}{\pi \, \nu}\, \tom_{2k-1,1-2j},
\end{split}
\end{equation}
with 
\begin{equation} \label{omegatilde}
\begin{split}
\tilde{\omega}_{2k-1,1-2j}=\! \! 
\sum\limits_{\alpha=\pm} \, 
\int\limits_{-\infty}^\infty dx \, e^{(2k-1)(x+i \alpha \, \eta_\alpha)} \,
{\cal F}_\alpha(x+i \, \alpha \, \eta_\alpha) \, 
 {\cal G}_{1-2j}(x+i \, \alpha \, \eta_\alpha), 
\quad j,k=1,2,...
\end{split}
\end{equation}
such that, ${\cal G}_{1-2j}(x)$ is the solution of the linear equation:
\begin{equation} \label{G1m2j}
\begin{split}
{\cal G}_{1-2j}(x)\!-\!\sum\limits_{\alpha=\pm}  
\int\limits_{-\infty}^\infty \frac{dy}{2 \pi} \, G(x\!-\!~y\!-\! i \,\alpha \eta_\alpha) \,
{\cal F}_\alpha(y+i \, \alpha \, \eta_\alpha) 
\, {\cal G}_{1-2j}(y+i \, \alpha \, \eta_\alpha)=e^{(1-2j)x},
\end{split}
\end{equation}
with $G(x)$ being the  kernel (\ref{G}) of the NLIE. 

To close this part, we emphasize again the fact that in the presence of a chemical potential, 
the singularities of $L_\pm(\theta)$ or equivalently ${\cal F}_\pm(\theta)$ deviate from 
the real axis. Consequently, the contour deformation parameters $\eta_\pm$ cannot be tend to zero, 
as in the zero chemical potential case of refs. \cite{Destri:1994bv} and \cite{Jimbo:2010jv}.


\section{Summary} \label{sect6}

In this paper, we derive the Kl\"umper-Batchelor-Pearce-Destri-de Vega
type nonlinear integral equation (NLIE) description of the thermodynamics of the 
sine-Gordon/massive Thirring model in the presence of a chemical potential 
coupled to the topological charge. The derivation employs the so-called Euclidean trick to relate the free energy to the ground state energy of the mirror-rotated theory, which turns out to be equivalent to a massive Thirring model with chemical-potential-dependent twisted boundary conditions (\ref{twistedBC}).

We further extend the equations to include an additional contribution to the thermodynamic potential associated with momentum conservation (\ref{DDVcont1}).

A key advantage of the resulting equations is their validity at arbitrary coupling and their efficiency in computing expectation values of local operators and certain currents as functions of temperature and chemical potential.

This description is particularly valuable for two reasons. First, it provides a testing ground for alternative approaches, such as the method of random surfaces \cite{Toth:2024lrv}. Second, by enabling efficient computation of vertex operator expectation values, it offers a theoretical prediction of the experimentally measurable coherence factors in tunnel-coupled cold atomic condensates \cite{Hofferberth:2007qfs,Esslernincs}.

\vspace{1cm}
{\tt Acknowledgments}

\noindent 
The authors thank J\'anos Balog and Zolt\'an Bajnok for useful discussions.
This research was supported  by the NKFIH grant K134946.

\appendix

\section{Some numerical results} \label{appA}

This short appendix is to provide with some numerical data  which were used to test the 
NLIE (\ref{DDVcont1}) agains the TBA equations (\ref{TBAeqs}). In the columns for the energy 
and momentum, only one number can be found, because using appropriately dense distribution 
for the discretization points, the TBA and NLIE results do not deviate within machine precision. 
So, we found it pointless to write down the same number twice. 


\begin{table}[h]
\begin{center}
\begin{tabular}{|c|c|c|c|}
\hline
$\beta$ & $ \mu \beta $   & $ -6\, \beta \, E_0^{(\beta,\mu)}/\pi $ & $c_{eff}$   \tabularnewline
 \hline
 $1$  & $0.12$ & $0.739896099595254 $ &   $1$\\
 \hline
$10^{-1}$  & $1.2 \cdot 10^{-2}$ & $0.995722741309564 $ & $1$  \\
 \hline
$10^{-2}$  & $1.2 \cdot 10^{-3}$ & $0.999953554086839 $  & $1$ \\
 \hline
$10^{-3}$  & $1.2 \cdot 10^{-4}$ & $0.999999527626539 $ &  $1$ \\
 \hline
$10^{-4}$  & $1.2 \cdot 10^{-5}$ & $0.999999995259212 $  & $1$ \\
 \hline
\end{tabular}\label{table_1}
\bigskip
\caption{Numerical data for the effective central charge  at $p=1/2,$ when the $\mu$ is kept fixed. 
The last column contains the prediction to the conformal limit given in (\ref{Econfc1}). }
\label{text1}
\end{center}
\end{table}
\normalsize

\begin{table}[h]
\begin{center}
\begin{tabular}{|c|c|c|c|}
\hline
$\beta$ & $ \mu \, \beta $   & $ -6\, \beta \, E_0^{(\beta,\mu)}/\pi $ & $c_{eff}$   \tabularnewline
 \hline
 $1$  & $0.12$ & $0.739896099595254$ &   $1.002918050088899$\\
 \hline
$10^{-1}$  & $0.12$ & $0.998611083334782 $ & $1.002918050088899$  \\
 \hline
$10^{-2}$  & $0.12$ & $1.002871311222578 $  & $1.002918050088899$ \\
 \hline
$10^{-3}$  & $0.12$ & $1.002917574794913 $ &  $1.002918050088899$ \\
 \hline
$10^{-4}$  & $0.12$ & $1.002918045318927 $  & $1.002918050088899$ \\
 \hline
\end{tabular}\label{table_2}
\bigskip
\caption{Numerical data for the effective central charge  at $p=1/2,$ when the $\mu/T$ is kept fixed. 
The last column contains the prediction to the conformal limit given in (\ref{Econfmu0}). }
\label{text2}
\end{center}
\end{table}
\normalsize

\begin{table}[h]
\begin{center}
\begin{tabular}{|c|c|c|c|c|}
\hline
$\beta$ & $ \mu  $ & $\lambda$  & $  E_0^{(\beta,\mu)}$ & $  P_0^{(\beta,\mu)}$  \tabularnewline
 \hline
 $1$  & $0.12$ &  $0.69$ & $-0.843492398573790 $ &   $0.582009755015915 $\\
 \hline
$10^{-1}$  & $0.12$ & $0.69$ & $-9.971518321712213 $ & $6.880347641981426 $  \\
 \hline
$10^{-2}$  & $0.12$ & $0.69$ & $-99.940075198159632 $  & $68.958651886730137 $ \\
 \hline
$10^{-3}$  & $0.12$ & $0.69$ & $-999.424788379081633 $ &  $689.603103981566277 $ \\
 \hline
$10^{-4}$  & $0.12$ & $0.69$ & $-9994.250320767618177 $  & $6896.032721327733270 $ \\
 \hline
\end{tabular}\label{table_3}
\bigskip
\caption{Numerical data for the energy and momentum  at $p=1/2,$ when the momentum is also 
included in the thermodynamic potential.  }
\label{text3}
\end{center}
\end{table}
\normalsize




\clearpage




\newpage


\begin{thebibliography}{99}



\bibitem{KlumperPearce}
A. Kl\"umper and M. T. Batchelor and P A Pearce,
”Central charges of the 6- and 19-vertex models with twisted boundary conditions”,
Journal of Physics A: Mathematical and General \textbf{24} (1991) 3111-3133.
doi:10.1088/0305-4470/24/13/025

\bibitem{Destri:1992qk}
C.~Destri and H.~J.~de Vega, 
``New thermodynamic Bethe ansatz equations without strings,''
Phys. Rev. Lett. \textbf{69} (1992), 2313-2317 
doi:10.1103/PhysRevLett.69.2313 

\bibitem{Klumper:1993vq}
A.~Kl\"umper,
``Thermodynamics of the Anisotropic Spin-1/2 Heisenberg Chain and Related Quantum Chains,''
Z. Phys. B \textbf{91} (1993), 507
doi:10.1007/BF01316831
[arXiv:cond-mat/9306019 [cond-mat]].

\bibitem{Destri:1994bv}
C.~Destri and H.~J.~De Vega, 
``Unified approach to thermodynamic Bethe Ansatz and finite size corrections for lattice models and field theories,'' 
Nucl. Phys. B \textbf{438} (1995), 413-454 
doi:10.1016/0550-3213(94)00547-R 
[arXiv:hep-th/9407117 [hep-th]].


\bibitem{FW2}
 M. Fowler and X. Zotos,
 ``Bethe-ansatz quantum sine-Gordon thermodynamics. The specific heat,''
{\em Phys. Rev. }{\bf 
B25} (1982) 5806.


\bibitem{Nagy:2023phz}
B.~C.~Nagy, G.~Tak{\'a}cs and M.~Kormos,
``Thermodynamic Bethe Ansatz and generalised hydrodynamics in the sine-Gordon model,''
SciPost Phys. \textbf{16} (2024) no.6, 145
doi:10.21468/SciPostPhys.16.6.145
[arXiv:2312.03909 [cond-mat.str-el]].



\bibitem{Essler2005}
Fabian H.~L. Essler, Holger Frahm, Frank G{\"o}hmann, Andreas Kl{\"u}mper, and Vladimir E. Korepin.
\textit{The One-Dimensional Hubbard Model}. 
Cambridge University Press, 2005. online ISBN: 9780511534843. \\
doi: \href{https://doi.org/10.1017/CBO9780511534843}{10.1017/CBO9780511534843}.

\bibitem{Takahashi:1972zza}
M.~Takahashi and M.~Suzuki,
``One-Dimensional Anisotropic Heisenberg Model at Finite Temperatures,''
Prog. Theor. Phys. \textbf{48} (1972), 2187-2209
doi:10.1143/PTP.48.2187

\bibitem{s-coleman}S. Coleman, ``The Quantum Sine-Gordon Equation as the Massive Thirring Model,''
\emph{Phys. Rev.} \textbf{D11} (1975) 2088.



\bibitem{Klassen:1992eq}
T.~R.~Klassen and E.~Melzer,
``Sine-Gordon not equal to massive Thirring, and related heresies,''
Int. J. Mod. Phys. A \textbf{8} (1993), 4131-4174 
doi:10.1142/S0217751X93001703 
[arXiv:hep-th/9206114 [hep-th]]. 
 
 
\bibitem{Feverati:2000xa} 
G.~Feverati, 
``Finite volume spectrum of Sine-Gordon model and its restrictions,'' 
[arXiv:hep-th/0001172 [hep-th]]. 



\bibitem{Zamolodchikov:1989cf}
A.~B.~Zamolodchikov, 
``Thermodynamic Bethe Ansatz in Relativistic Models. Scaling Three State Potts and Lee-yang Models,'' 
Nucl. Phys. B \textbf{342} (1990), 695-720 
doi:10.1016/0550-3213(90)90333-9 

\bibitem{ddvlc} C. Destri and H.J. de Vega, ``Light-cone lattice approach to fermionic theories in 2D,''
\emph{Nucl. Phys.} \textbf{B290} (1987) 363-391.


\bibitem{Hegedus:2017muz}
{\'A}.~Heged{\'{u}}s,
``Lattice approach to finite volume form-factors of the Massive Thirring (Sine-Gordon) model,''
JHEP \textbf{08} (2017), 059
doi:10.1007/JHEP08(2017)059
[arXiv:1705.00319 [hep-th]].


\bibitem{Zamolodchikov:1991et}
A.~B.~Zamolodchikov,
``On the thermodynamic Bethe ansatz equations for reflectionless ADE scattering theories,''
Phys. Lett. B \textbf{253} (1991), 391-394
doi:10.1016/0370-2693(91)91737-G


\bibitem{Zamolodchikov:1978xm} 
A.~B.~Zamolodchikov and A.~B.~Zamolodchikov, 
``Factorized s Matrices in Two-Dimensions as the Exact Solutions of Certain Relativistic Quantum Field Models,''
Annals Phys. \textbf{120} (1979), 253-291 
doi:10.1016/0003-4916(79)90391-9 


\bibitem{Castro-Alvaredo:2016cdj}
O.~A.~Castro-Alvaredo, B.~Doyon and T.~Yoshimura,
``Emergent hydrodynamics in integrable quantum systems out of equilibrium,''
Phys. Rev. X \textbf{6} (2016) no.4, 041065
doi:10.1103/PhysRevX.6.041065
[arXiv:1605.07331 [cond-mat.stat-mech]].


\bibitem{Doyon:2019nhl}
B.~Doyon,
``Lecture notes on Generalised Hydrodynamics,''
SciPost Phys. Lect. Notes \textbf{18} (2020), 1
doi:10.21468/SciPostPhysLectNotes.18
[arXiv:1912.08496 [cond-mat.stat-mech]].


\bibitem{Jimbo:2010jv}
M.~Jimbo, T.~Miwa and F.~Smirnov,
``Hidden Grassmann structure in the XXZ model V: Sine-Gordon model,''
Lett. Math. Phys. \textbf{96} (2011), 325-365 
doi:10.1007/s11005-010-0438-9 
[arXiv:1007.0556 [hep-th]]. 


\bibitem{Boos:2006mq}
H.~Boos, M.~Jimbo, T.~Miwa, F.~Smirnov and Y.~Takeyama, 
``Hidden Grassmann structure in the XXZ model,'' 
Commun. Math. Phys. \textbf{272} (2007), 263-281 
doi:10.1007/s00220-007-0202-x 
[arXiv:hep-th/0606280 [hep-th]].

\bibitem{Boos:2008rh} 
H.~Boos, M.~Jimbo, T.~Miwa, F.~Smirnov and Y.~Takeyama, 
``Hidden Grassmann Structure in the XXZ Model II: Creation Operators,''
Commun. Math. Phys. \textbf{286} (2009), 875-932 
doi:10.1007/s00220-008-0617-z 
[arXiv:0801.1176 [hep-th]]. 

\bibitem{Jimbo:2008kn} 
M.~Jimbo, T.~Miwa and F.~Smirnov, 
``Hidden Grassmann Structure in the XXZ Model III: Introducing Matsubara direction,''
J. Phys. A \textbf{42} (2009), 304018 
doi:10.1088/1751-8113/42/30/304018 
[arXiv:0811.0439 [math-ph]]. 

\bibitem{Boos:2010qii}
H.~Boos, M.~Jimbo, T.~Miwa and F.~Smirnov, 
``Hidden Grassmann Structure in the XXZ Model IV: CFT limit,''
Commun. Math. Phys. \textbf{299} (2010), 825-866 
doi:10.1007/s00220-010-1051-6 
[arXiv:0911.3731 [hep-th]]. 


\bibitem{Jimbo:2010jv}
M.~Jimbo, T.~Miwa and F.~Smirnov,
``Hidden Grassmann structure in the XXZ model V: Sine-Gordon model,''
Lett. Math. Phys. \textbf{96} (2011), 325-365 
doi:10.1007/s11005-010-0438-9 
[arXiv:1007.0556 [hep-th]]. 


\bibitem{Hegedus:2019rju} 
\'A.~Heged\'{u}s, 
``Finite volume expectation values in the sine-Gordon model,'' 
JHEP \textbf{01} (2020), 122 
doi:10.1007/JHEP01(2020)122 
[arXiv:1909.08467 [hep-th]]. 


\bibitem{Zamolodchikov:1995xk} 
A.~B.~Zamolodchikov, 
``Mass scale in the sine-Gordon model and its reductions,'' 
Int. J. Mod. Phys. A \textbf{10} (1995), 1125-1150 
doi:10.1142/S0217751X9500053X 


\bibitem{Toth:2024lrv}
M.~T{\'o}th, J.~H.~Pixley, D.~Sz{\'a}sz-Schagrin, G.~Tak{\'a}cs and M.~Kormos,
``Sine-Gordon model at finite temperature: The method of random surfaces,''
Phys. Rev. B \textbf{111} (2025) no.15, 155112
doi:10.1103/PhysRevB.111.155112
[arXiv:2408.08828 [cond-mat.stat-mech]].


\bibitem{Bajnok:2000wm}
Z.~Bajnok, L.~Palla, G.~Takacs and F.~Wagner,
``The k folded sine-Gordon model in finite volume,''
Nucl. Phys. B \textbf{587} (2000), 585-618
doi:10.1016/S0550-3213(00)00441-7
[arXiv:hep-th/0004181 [hep-th]].

\bibitem{Ahn:2011xq}
C.~Ahn, Z.~Bajnok, D.~Bombardelli and R.~I.~Nepomechie,
``TBA, NLO Luscher correction, and double wrapping in twisted AdS/CFT,''
JHEP \textbf{12} (2011), 059
doi:10.1007/JHEP12(2011)059
[arXiv:1108.4914 [hep-th]].


\bibitem{Bajnok:2007jg}
Z.~Bajnok and Z.~Simon,
``Solving topological defects via fusion,''
Nucl. Phys. B \textbf{802} (2008), 307-329
doi:10.1016/j.nuclphysb.2008.04.003
[arXiv:0712.4292 [hep-th]].




\bibitem{Feverati:1999sr}
G.~Feverati, F.~Ravanini and G.~Takacs,
``Nonlinear integral equation and finite volume spectrum of minimal models perturbed by $\phi_{(1,3)}$,''
Nucl. Phys. B \textbf{570} (2000), 615-643
doi:10.1016/S0550-3213(99)00771-3
[arXiv:hep-th/9909031 [hep-th]].

\bibitem{Fioravanti:1996rz}
D.~Fioravanti, A.~Mariottini, E.~Quattrini and F.~Ravanini,
``Excited state Destri-De Vega equation for Sine-Gordon and restricted Sine-Gordon models,''
Phys. Lett. B \textbf{390} (1997), 243-251
doi:10.1016/S0370-2693(96)01409-8
[arXiv:hep-th/9608091 [hep-th]].






\bibitem{Fendley:1993jh}
P.~Fendley and H.~Saleur,
``Massless integrable quantum field theories and massless scattering in (1+1)-dimensions,''
[arXiv:hep-th/9310058 [hep-th]].



\bibitem{Hofferberth:2007qfs}
S.~Hofferberth, I.~Lesanovsky, B.~Fischer, T.~Schumm and J.~Schmiedmayer,
``Non-equilibrium coherence dynamics in one-dimensional Bose gases,''
Nature \textbf{449} (2007) no.7160, 324-327
doi:10.1038/nature06149



\bibitem{Esslernincs}
Y. D. van Nieuwkerk, J. Schmiedmayer, F.~H.~L.~Essler,
``Projective phase measurements in one-dimensional Bose gases,''
SciPost Phys. \textbf{5} (2018) no.5, 046
https://doi.org/10.21468/SciPostPhys.5.5.046
[arXiv: arXiv:1806.02626 [cond-mat.str-el]].



\end{thebibliography}
\end{document}